**Title: Room temperature nanocavity laser with interlayer excitons in 2D heterostructures**

**Authors**

Yuanda Liu,[1,2†] Hanlin Fang,[3†] Abdullah Rasmita,[1] Yu Zhou,[1] Juntao Li,[3] Ting Yu,[1] Qihua Xiong,[1] Nikolay Zheludev,[1,2,4*] Jin Liu,[3*] and Weibo Gao[1,2*]

**Affiliations**

[1]Division of Physics and Applied Physics, School of Physical and Mathematical Sciences, Nanyang Technological University, Singapore. [2]The Photonics Institute and Centre for Disruptive Photonic Technologies, Nanyang Technological University, Singapore. [3]State Key Laboratory of Optoelectronic Materials and Technologies, School of Physics, Sun Yat-sen University, Guangzhou 510275, China. [4]Optoelectronics Research Centre, University of Southampton, UK.

†These authors contributed equally to this work.
*Corresponding authors. Email: nzheludev@ntu.edu.sg (N.Z.); liujin23@mail.sysu.edu.cn (J.L.); wbgao@ntu.edu.sg (W.G.)

**Abstract**

Atomically thin layered two dimensional (2D) material has provided a rich library for both fundamental research and device applications. One of the special advantages is that, bandgap engineering and controlled material response can be achieved by stacking different 2D materials. Recently several types of excitonic lasers have been reported based on Transition metal dichalcogenide (TMDC) monolayers, however, the emission is still the intrinsic energy bandgap of the monolayers and lasers harnessing the flexibility of Van der Waals heterostructures have not been demonstrated yet. Here, we report for the first time a room temperature interlayer exciton laser with $MoS_2/WSe_2$ heterostructures. The onset of lasing action was identified by a combination of distinct kink in the 'L-L' curve and the noticeable collapse of spectral linewidth. Different from visible emission of intralayer excitons for both $MoS_2$ and $WSe_2$, our interlayer exciton laser works in the infrared range, which is fully compatible with the well-established technologies in silicon photonics. Thanks to the long lifetime of interlayer excitons, the requirement of the cavity quality factor is relaxed by orders of magnitude. The demonstration of room temperature interlayer exciton laser might open new perspectives for the development of coherent light source with tailored optical properties on silicon photonics platform.

**MAIN TEXT**

**Introduction**

Investigations of 2D layered materials have attracted substantial attention in recent years due to their novel electronic and optical properties as compared to traditional 3D bulk materials(*1*). The surface of 2D layers is free of dangling bonds and different layers of 2D materials can be stacked together to form heterostructures with Van der Waals forces(*2, 3*). This largely eases the conventional 'lattice mismatch' issue and provides even richer platforms since the properties of these Van der Waals heterostructures can be engineered by the material types, lattice alignment and coupling strengths of different layers.

Among them, one particular type is the heterostructures with transition-metal dichalcogenide (TMDC) materials. Unlike graphene, monolayer TMDCs are direct bandgap material(*4, 5*) and their excitons can be used for light-emitting applications. To list only a few, light emitting diodes(*6-8*), photon detectors(*9*), field effect transistors(*10*) and opto-valleytronic devices(*11*) have been demonstrated. Furthermore, quantum yield for photoluminescence (PL) of TMDCs allows them to act as gain medium to produce monolayer excitonic lasers, as demonstrated in several pioneer works recently(*12-15*). Different from monolayer TMDC intralayer excitons, heterostructures formed by these materials can emit light with interlayer excitons(*16, 17*), with electron and holes separated in different TMDC layers(*18*). Indirect exciton in TMDC heterostructures represents one type of bandgap engineering and the emission can cover a more flexible wavelength range.

Here we demonstrate the first nanocavity laser based on Van der Waals heterostructures. Figure 1a presents a three-dimensional schematic of our nanolaser device and Figure 1b shows the optical microscope image of the real device. Van der Waals $MoS_2/WSe_2$ heterostructure has been chosen to play the role of the gain medium on a vertically coupled, free-standing photonic crystal cavity (PhCC). We have specially designed the cavity structure such that the field maximum is in excellent spatial overlap with the gain medium (Details shown in Note S1). The designed nanocavity facilitates efficient funneling of spontaneous emission from the Van der Waals gain material into the cavity mode.

**Results**

$WSe_2$ and $MoS_2$ monolayers were mechanically exfoliated from bulk single crystals onto the polydimethylsiloxane (PDMS) stamps, followed by stacking onto the cavity consecutively with dry transfer method. The monolayers were identified and selected by using a combination of optical contrast (Figure 1b), photoluminescence (PL) and Raman spectroscopy (Fig. S1). As shown in Figure 1c, the intralayer exciton recombination in $MoS_2$ and $WSe_2$ monolayers results in PL peaks at about 668 nm and 750 nm, respectively. For $MoS_2/WSe_2$ heterostructures with AA stacking, a new PL peak centering at 1,128.6 nm has been observed (Figure 1c), which we attribute to the indirect excitonic emission. Several different samples have been measured to confirm the indirect exciton energy, which agrees well with previously reported results measured with scanning tunneling spectroscopy(*19*). As shown in Figure 1d, when indirect exciton is coupled to the cavity, a sharp cavity mode in the emission spectrum was observed with wavelength located at ~ 1,122 nm, corresponding to the designed cavity mode of the PhCC. From the figure 1d, we can also see that, the silicon emission wavelength (1097.5 nm) is shorter than that of indirect excitons, implying silicon has a larger bandgap, leading to a minimum absorption of indirect exciton emission, and therefore serves as an ideal cavity material for our infrared laser.

Next, we proceed to prove the lasing behavior of our device. One of the laser properties is that emission intensity in the cavity mode will increase faster when the pump power is above the lasing threshold. First, we investigate the lasing behavior at cryo-temperature ~ 5K. Figure 2a presents the evolution of the PL spectrum as a function of the excitation power around the lasing threshold. In order to see the spontaneous emission more clearly, here we collect with a multimode fiber in contrast to Figure 1d of single-mode fiber collection. The broad emission spectrum corresponds to the spontaneous emission of indirect excitons. With the increase of the excitation power, the cavity mode becomes more apparent as compared to the background. The experimental data can be Lorentzian fitted with a lasing peak component and a spontaneous emission component, as shown in Figure 2b. Figure 2c exhibits their emission intensity as a function of excitation power, which is referred to as the 'light input-light output' or 'L-L' curve. When the excitation intensity exceeds a certain value (threshold), we can observe the rapid increase of cavity emission intensity, indicating that stimulated emission starts taking place in our device. In contrast, no kink signature was observed from the power-dependent data for spontaneous emission.

Another figure of merit for lasing is the linewidth narrowing behavior. The inset in Figure 2d exhibits that the spectral linewidth of the cavity mode decreases gradually as the excitation power increases. The linear plot of linewidth (FWHM, full-width at half-maximum) values as a function of the pump power displays a pronounced kink and plateau in the linewidth trace. The linewidth narrows by ~ 7%, that is, from ~2.85 nm to ~2.65 nm.

More evidence of lasing can be obtained from the coherence time vs. pump power measurement, the coherent length was measured using a free-beam Michelson interferometer as shown in Supplementary Figure 2. The retroreflector was mounted on a linear stage to generate delay between two optical paths. The typical result of this measurement for the input power of 250 μW is shown in Figure 3. As shown in Figure 3b, based on the raw data of intensity vs. path delay, we extract the visibility as a function of path delay (Figure 3b). Here the visibility is defined as $\frac{I_{max}(\Delta\tau) - I_{min}(\Delta\tau)}{I_{max}(\Delta\tau) + I_{min}(\Delta\tau)}$ where $I_{max(min)}(\Delta\tau)$ is the maximum (minimum) intensity of the envelope function at a delay $\Delta\tau$. We can see that the visibility contains two components: a broad component from the lasing mode contribution, and a narrow component from the non-lasing mode contribution. Two peaks Gaussian fitting is used to extract these two contributions. The coherence time of the lasing mode was extracted from the full width at half maximum (FWHM) of the visibility vs path delay data. Pump power dependence of coherence time is shown in Figure 3c. It can be seen that the coherence time reaches a saturation limit of 1.7 ps at ~35 μW, which is closed to the lasing threshold obtained from the 'L-L' curve. It is worth noting that the saturation limit of 1.7 ps is agreeable with the coherence time derived from the laser linewidth (~ 1.9 ps).

The rather soft turn-on shoulder near the threshold implies a large β factor, which is defined as the ratio of the spontaneous emission coupled into the laser cavity mode over the total spontaneous emission from the material. Therefore, β factor is a measure of the optical efficiency of a laser. Its theoretical limit is unity, corresponding to the case of thresholdless lasing. We estimated the β value by fitting log-log plot of the L-L curve to the theoretical graph based on classical laser rate equations (Figure 2e). The solid red curve is the best fit to the experimental data, yielding a β value of 0.17. Figure 2e also shows theoretical curves of β = 0.1, β = 0.4, and β = 1 as a reference. Based on the fitting of the L-L curve to the rate equation model, we extract a lasing threshold of ~ 33 μW. More detail of the theoretical model and fitting method can be found in Note S4 and Note S5.

The quality (Q) factor was estimated from Q = λ/Δλ, where Δλ is the FWHM (~ 2.65 nm) of the spectral cavity emission peak just above the threshold, to be ~ 423. It is worth mentioning that interlayer excitons have a much longer lifetime ~ ns as compared to the intralayer exciton ~ ps. This actually largely relaxes the cavity requirement for the lasing. Indeed, our cavity Q factor above is much lower than the previously reported intralayer exciton lasers(*12-15*), but lasing can still happen here. This is mainly because of the longer lifetime of the interlayer exciton. It can be shown that for the same cavity quality, longer carrier lifetime will result in a lower lasing threshold. We have included more detailed discussion in Note S5.

Thanks to the large exciton binding energy in 2D materials, high-performance lasing behavior of our device is well sustained even at room temperature. Figure 4a shows the lasing spectrum and the spontaneous emission measured at room temperature (295 K) when the excitation spot moves on and away from the cavity. The first mode peak is located at ~1,128 nm, which is redshifted by about 8 nm comparing with that measured at 5 K. This comes from the energy shift of the cavity mode with varying the sample temperature. Similar to the low-temperature case, a kink in the L-L curve and linewidth narrowing is observed (Fig. 4b, 4c). The FWHM of the cavity mode slightly above the threshold is ~ 2.15 nm (Figure 3c), which is slightly lower than that in 5K and represents an increase of the quality factor of the cavity in room temperature. Figure 4d exhibits the log-log plot of the room-temperature cavity mode intensity as a function of the pump power. The solid lines represent the simulations using standard laser rate equations. The best fit gives rise to a β factor of 0.27, signifying 10% higher coupling efficiency than that at 5 K. Lasing threshold at room temperature is estimated to be ~54 μW.

**Discussion**

In summary, for the first time, we have realized a room temperature near-infrared laser based on interlayer excitons in $MoS_2$/$WSe_2$ van der Waals heterostructures. The lasing emission was manifested by a kink feature in the 'L-L' curve, spectral linewidth narrowing. The flexibility of heterostructures brings more possibilities in the future(*20*). Ternary phase layered material such as $MoS_xSe_y$ or $WS_xSe_y$ has been demonstrated to have a tunable wavelength(*21, 22*). Vertical stacking of these materials could engineer more possible bandgap. The wide range of 2D materials, combined with stacking orientations, different layer numbers, allows for the controlled material response and shows a great promise for their nanophotonics applications. Future full integration of coherent light sources, waveguides, modulators and detectors on single chips may eventually lead to on-chip 2D material integrated photonics(*1, 23*).

**Materials and Methods**
**Cavity design**

The PhCC was fabricated from an SOI wafer comprising a 220 nm thick silicon layer on a 2 μm sacrificial silicon dioxide layer. The mask pattern was defined by a standard electron-beam lithography process and then transferred into the silicon membrane using an HBr-based inductively coupled plasma RIE. The silica beneath the PhCC was finally removed using a hydrofluoric acid solution to provide vertical optical isolation. This fabrication process forms an air-bridged PhCC slab with an air hole array, which produces an in-plane photonic bandgap.

**Heterobilayer fabrication**

$WSe_2$ and $MoS_2$ monolayers were mechanically exfoliated from bulk single crystals onto the polydimethylsiloxane (PDMS) stamps, followed by stacking onto the cavity consecutively in two dry transfer steps. The monolayers were identified and selected by using a combination of optical contrast, photoluminescence and Raman spectroscope. The heterobilayer was subsequently thermally annealed in a flow of forming gas (Ar:$H_2$ = 95:5 by volume) at 200 ºC for 3 hours.

**Optical characterization**

Samples were mounted on the cold head of a closed-cycle helium cryostat with variable temperature (5 K < T < 350 K). For steady-state micro-photoluminescence measurement, we excited the system with an above-bandgap (λ = 740 nm) continuous wave diode laser focused to a 1 μm spot on the surface of the sample using a 50 × long working distance infrared microscope objective lens (Numerical Aperture = 0.65), and positioned the laser spot onto the PhCC using piezo-electric nanopositioners. The photoluminescence signal was collected by the same objective, dispersed by a 300 $mm^{-1}$ grating monochromator with a spectral resolution of 0.16 nm, and detected by a liquid-nitrogen-cooled charge coupled device (CCD) camera. The pumping power was adjusted using a neutral density filter.

**Supplementary Materials**

Note S1. Photonic crystal cavity design
Note S2. Raman spectra of monolayer $MoS_2$ and $WSe_2$
Note S3. Coherence time measurement setup
Note S4. Beta factor fitting
Note S5. Effect of the lifetime on lasing threshold
Fig. S1. Raman spectra of monolayer $MoS_2$ and $WSe_2$
Fig. S2. Michelson interferometer for coherent length measurement.
Fig. S3. Illustration of the light-matter interaction.
Fig. S4. Lasing threshold vs. exciton lifetime.
Fig. S5. Comparison between the PL spectrum from $MoS_2$/$WSe_2$ heterostructure and the PL spectrum from the silicon substrate.
Fig. S6. Photonic cavity characterization.
Table S1. Definition and values of the parameters used in the theoretical model

**References and Notes**

1. F. Xia, H. Wang, D. Xiao, M. Dubey, A. Ramasubramaniam, Two-dimensional material nanophotonics. *Nat. Photonics* **8**, 899-907 (2014).
2. A. Geim, I. Grigorieva, Van der Waals heterostructures. *Nature* **499**, 419-425 (2013).
3. K. S. Novoselov, A. Mishchenko, A. Carvalho, A. H. Castro Neto, 2D materials and van der Waals heterostructures. *Science* **353**, (2016).
4. K. F. Mak, C. Lee, J. Hone, J. Shan, T. F. Heinz, Atomically thin $MoS_2$: a new direct-gap semiconductor. *Phys. Rev. Lett.* **105**, 136805 (2010).
5. A. Splendiani, L. Sun, Y. Zhang, T. Li, J. Kim, C.-Y. Chim, G. Galli, F. Wang, Emerging photoluminescence in monolayer $MoS_2$. *Nano Lett.* **10**, 1271-1275 (2010).
6. R. S. Sundaram, M. Engel, A. Lombardo, R. Krupke, A. C. Ferrari, P. Avouris, M. Steiner, Electroluminescence in single layer $MoS_2$. *Nano Lett.* **13**, 1416-1421 (2013).
7. B. W. H. Baugher, H. O. H. Churchill, Y. Yang, P. Jarillo Herrero, Optoelectronic devices based on electrically tunable p-n diodes in a monolayer dichalcogenide. *Nat. Nanotechnol.* **9**, 262-267 (2014).
8. J. S. Ross, P. Klement, A. M. Jones, N. J. Ghimire, J. Yan, D. Mandrus, T. Taniguchi, K. Watanabe, K. Kitamura, W. Yao, Electrically tunable excitonic light-emitting diodes based on monolayer $WSe_2$ p-n junctions. *Nat. Nanotechnol.* **9**, 268-272 (2014).
9. O. Lopez-Sanchez, D. Lembke, M. Kayci, A. Radenovic, A. Kis, Ultrasensitive photodetectors based on monolayer $MoS_2$. *Nat. Nanotechnol.* **8**, 497-501 (2013).
10. B. Radisavljevic, A. Kis, Mobility engineering and a metal–insulator transition in monolayer MoS2. *Nat. Mater.* **12**, 815-820 (2013).
11. H. Yu, G.-B. Liu, P. Gong, X. Xu, W. Yao, Dirac cones and Dirac saddle points of bright excitons in monolayer transition metal dichalcogenides. *Nat. Commun.* **5**, 3876 (2014).
12. J. Shang, C. Cong, Z. Wang, N. Peimyoo, L. Wu, C. Zou, Y. Chen, X. Y. Chin, J. Wang, C. Soci, Room-temperature 2D semiconductor activated vertical-cavity surface-emitting lasers. *Nat. Commun.* **8**, 543 (2017).
13. Y. Li, J. Zhang, D. Huang, H. Sun, F. Fan, J. Feng, Z. Wang, C. Z. Ning, Room-temperature continuous-wave lasing from monolayer molybdenum ditelluride integrated with a silicon nanobeam cavity. *Nat. Nanotechnol.* **12**, 987 (2017).
14. Y. Ye, Z. J. Wong, X. Lu, X. Ni, H. Zhu, X. Chen, Y. Wang, X. Zhang, Monolayer excitonic laser. *Nat. Photonics* **9**, 733-737 (2015).

15. S. Wu, S. Buckley, J. R. Schaibley, L. Feng, J. Yan, D. G. Mandrus, F. Hatami, W. Yao, J. Vuckovic, A. Majumdar, Monolayer semiconductor nanocavity lasers with ultralow thresholds. *Nature* **520**, 69 (2015).
16. P. Rivera, J. R. Schaibley, A. M. Jones, J. S. Ross, S. Wu, G. Aivazian, P. Klement, K. Seyler, G. Clark, N. J. Ghimire, J. Yan, D. G. Mandrus, W. Yao, X. Xu, Observation of long-lived interlayer excitons in monolayer $MoSe_2$-$WSe_2$ heterostructures. *Nat. Commun.* **6**, 6242 (2015).
17. H. Fang, C. Battaglia, C. Carraro, S. Nemsak, B. Ozdol, J. S. Kang, H. A. Bechtel, S. B. Desai, F. Kronast, A. A. Unal, Strong interlayer coupling in van der Waals heterostructures built from single-layer chalcogenides. *Proc. Natl. Acad. Sci. U.S.A.* **111**, 6198-6202 (2014).
18. X. Hong, J. Kim, S.-F. Shi, Y. Zhang, C. Jin, Y. Sun, S. Tongay, J. Wu, Y. Zhang, F. Wang, Ultrafast charge transfer in atomically thin $MoS_2$/$WS_2$ heterostructures. *Nat. Nanotechnol.* **9**, 682-686 (2014).
19. C. Zhang, C.-P. Chuu, X. Ren, M.-Y. Li, L.-J. Li, C. Jin, M.-Y. Chou, C.-K. Shih, Interlayer couplings, Moiré patterns, and 2D electronic superlattices in $MoS_2$/$WSe_2$ hetero-bilayers. *Sci. Adv.* **3**, e1601459 (2017).
20. A. Pant, Z. Mutlu, D. Wickramaratne, H. Cai, R. K. Lake, C. Ozkan, S. Tongay, Fundamentals of lateral and vertical heterojunctions of atomically thin materials. *Nanoscale* **8**, 3870-3887 (2016).
21. S.-H. Su, W.-T. Hsu, C.-L. Hsu, C.-H. Chen, M.-H. Chiu, Y.-C. Lin, W.-H. Chang, K. Suenaga, H. He Jr, L.-J. Li, Controllable synthesis of band-gap-tunable and monolayer transition-metal dichalcogenide alloys. *Front. Energy Res.* **2**, 27 (2014).
22. A. F. Rigosi, H. M. Hill, K. T. Rim, G. W. Flynn, T. F. Heinz, Electronic band gaps and exciton binding energies in monolayer $Mo_xW_{1-x}S_2$ transition metal dichalcogenide alloys probed by scanning tunneling and optical spectroscopy. *Phys. Rev. B* **94**, 075440 (2016).
23. K. F. Mak, J. Shan, Photonics and optoelectronics of 2D semiconductor transition metal dichalcogenides. *Nat. Photonics* **10**, 216-226 (2016).
24. C. Gies, J. Wiersig, M. Lorke, F. Jahnke, Semiconductor model for quantum-dot-based microcavity lasers. *Phys. Rev. A* **75**, 013803 (2007).
25. G. Moody, C. K. Dass, K. Hao, C.-H. Chen, L.-J. Li, A. Singh, K. Tran, G. Clark, X. Xu, G. Berghäuser, Intrinsic homogeneous linewidth and broadening mechanisms of excitons in monolayer transition metal dichalcogenides. *Nat. Commun.* **6**, 8315 (2015).
26. M. P. van Exter, G. Nienhuis, J. P. Woerdman, Two simple expressions for the spontaneous emission factor β. *Phys. Rev. A* **54**, 3553-3558 (1996).
27. M. Kira, S. W. Koch, *Semiconductor quantum optics*. (Cambridge University Press, 2012).
28. K. Rivoire, A. Faraon, J. Vuckovic, Gallium phosphide photonic crystal nanocavities in the visible. *Appl. Phys. Lett.* **93**, 063103 (2008).

**Acknowledgments:**

**Funding:**

We acknowledge the support from the Singapore National Research Foundation through a Singapore 2015 NRF fellowship grant (NRF-NRFF2015-03) and its Competitive Research Program (CRP Award No. NRF-CRP14-2014-02), Singapore Ministry of Education (RG119/17(s); MOE2016-T2-2-077 and MOE2016-T3-1-006 (S)), AStar QTE programme, a start-up grant from NTU (M4081441),
the UK Engineering and Physical Sciences Research Council (EP/M009122/1). Jin Liu and Juntao Li thank the support from National Key R\&D Program of China (2018YFA0306100); National Natural Science Foundations of China (11874437,

**Author Contributions:** W.G. J. L. and N. Z. conceived and supervised the project. Y.L. Y. Z. performed experiments. F.L. J.L. J.L. fabricated cavity. A.R. conceived and carried out modeling. W. G, T. Y. and Q. X. together with other authors analyzed and discussed the results and co-wrote the manuscript.

**Competing interests:** The authors declare no competing interests.
**Data and materials availability:** All data needed to evaluate the conclusions in the paper are present in the paper and/or the Supplementary Materials. Additional data related to this paper may be requested from the authors.

# Figures

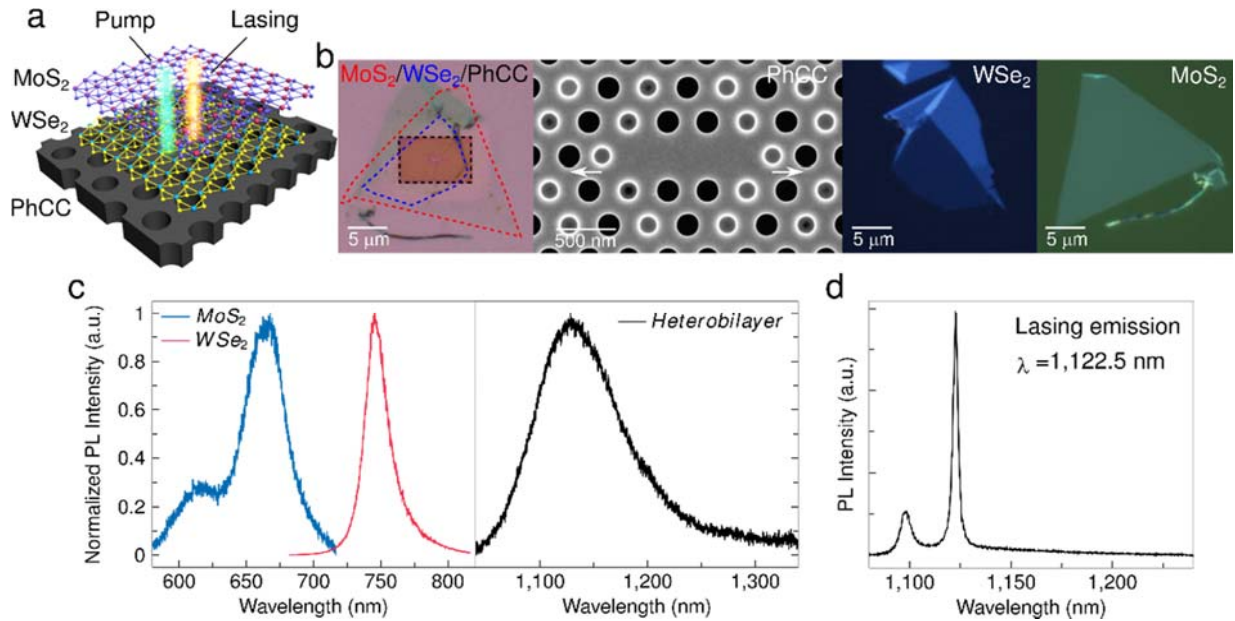

**Fig. 1. MoS$_2$/WSe$_2$ heterobilayer-PhCC nanolaser.** (**a**) 3D schematic image of the fabricated hetetobilayer-PhCC nanolaser. (**b**) First panel: Optical microscope image of MoS$_2$/WSe$_2$ heterobilayer on silicon PhCC. The MoS$_2$, WSe$_2$ monolayers, and PhCC are indicated by the red, blue and black dashed lines, respectively. Scale bar: 5 μm. Second panel: Top-view scanning electron micrograph around the defect region of the L3-type defect silicon photonic crystal cavity. Scale bar: 500 nm. The arrows show the displacement of the two end-holes by 0.163a (48.4 nm). The radius of other gray holes is shrunk to be ~62.4 nm to enhance the far-field vertical coupling. Third and fourth panels show the optical microscope image of WSe$_2$ and MoS$_2$ monolayers on PDMS before transfer, respectively. Scale bar: 5 μm. (**c**) Interlayer excitonic emission in MoS$_2$/WSe$_2$ heterobilayer, comparing with the intralayer emission in the two constituents. The spectra of monolayers were measured at room temperature with 532 nm excitation, and the spectrum of heterobilayer was measured at 5 K with 740 nm excitation. (**d**) Cavity lasing mode emission spectrum taken at a CW pump power of 190 μW, T = 5 K. Lasing action occurs at λ = 1,122.5 nm with a measured linewidth of ~2.7 nm.

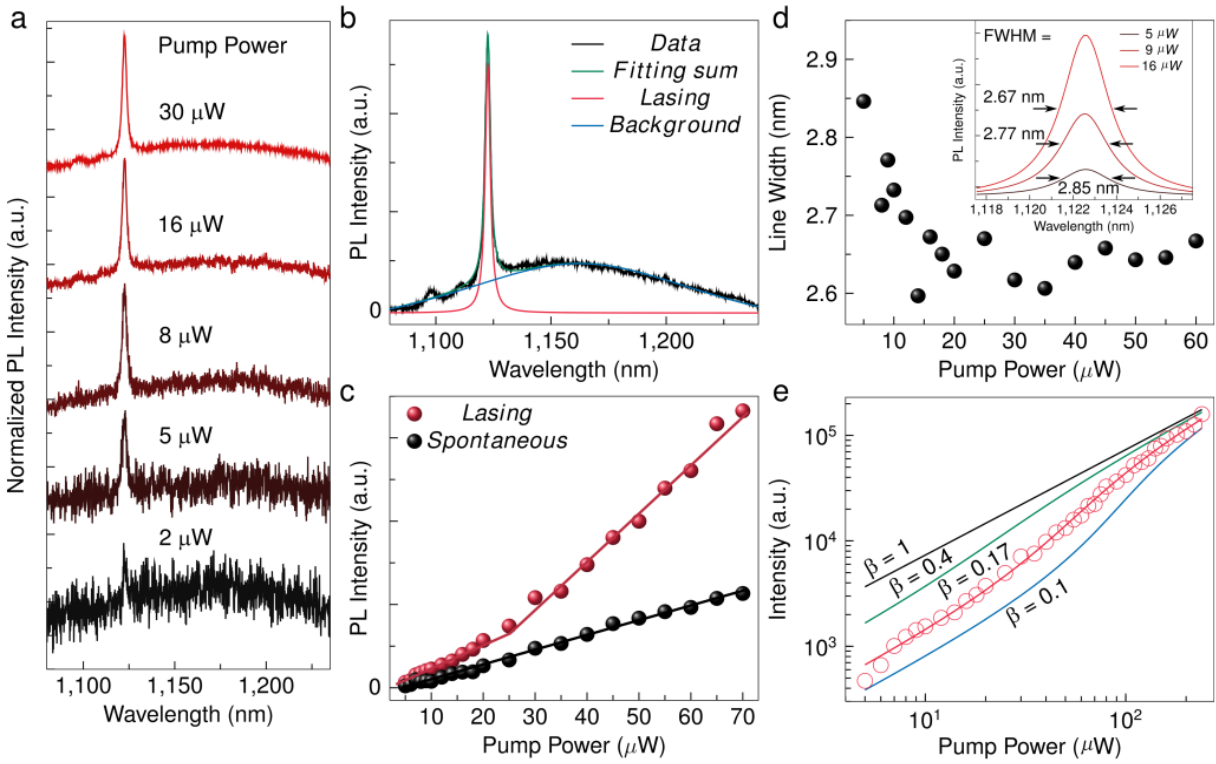

**Fig. 2. Indirect excitonic lasing characteristics.** (a) Steady-state PL emission spectra with increasing pump intensity around the threshold, illustrating the phase transition from spontaneous emission to stimulated emission. (b) Cavity lasing emission as compared to the heterobilayer photoluminescence background. Green line represents Lorentzian fits to the experimental data (black line), while red and blue lines represent lasing component and background component, respectively. (c) L-L curve showing the output intensity at the laser wavelength as a function of the excitation pump power. The cavity mode emission (red dots) exhibits a kink indicating the onset of superlinear emission and lasing operation, while the heterobilayer photoluminescence background emission (black dots) shows a linear dependence on the pump intensity. Solid lines are the linear fit to the experimental data. (d) The linewidth of the cavity mode emission as a function of the pump power. Inset: Cavity mode emission peaks with increasing pump power, exhibiting a linewidth narrowing behavior. (e) Log-Log plot of the cavity mode emission intensity as a function of the pump power. Solid lines are the simulated results of a rate-equation calculation with different β - factors. Red line shows the best fit to the experimental data (red circles) corresponding to β = 0.17. The fits of 0.1, 0.4, and 1 are also shown for comparison.

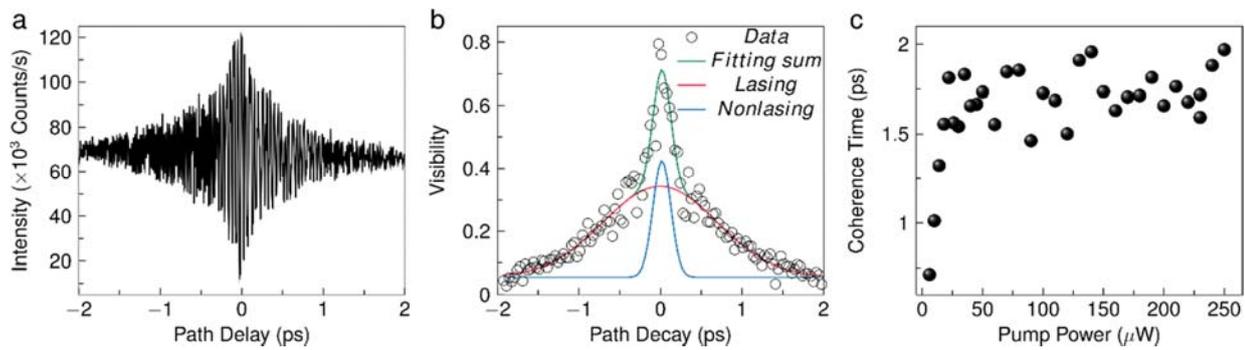

**Fig. 3. Coherence time measurement.** (**a**) Raw data of intensity as a function of path delay at pump power 250 μW. (**b**) Visibility as a function of path delay at pump power 250 μW. Two peaks Gaussian fitting is used to extract the contribution of lasing mode. (**c**) Coherence time as a function of pump power.

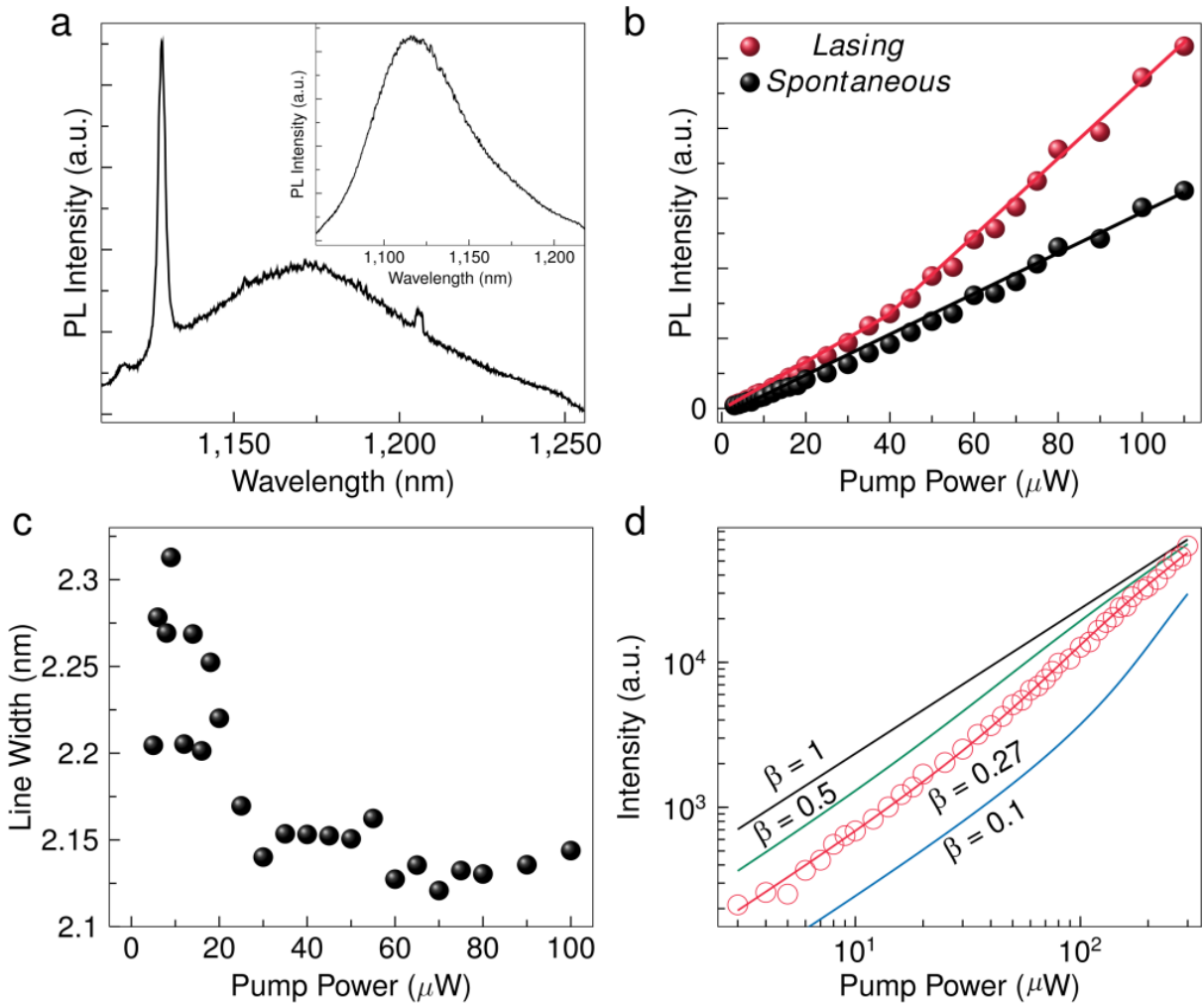

**Fig. 4. Room-temperature laser operation.** (**a**) Room temperature cavity lasing mode emission spectrum. The linewidth is ~2.26 nm. The inset shows the room-temperature off-cavity spectrum of the heterobilayer for comparison. (**b**) L-L curve showing the output intensity at the laser wavelength as a function of the excitation pump power. The cavity emission exhibits a kink indicating the onset of superlinear emission and lasing operation, while the background emission shows a linear dependence on the pump intensity. Red and black continuous lines are the linear fit to the experimental data. (**c**) The linewidth of the cavity mode emission as a function of the pump power, exhibiting linewidth narrowing profile near the threshold. (**d**) Log-Log plot of the cavity mode intensity as a function of the excitation pump power. Solid lines are the simulated results of a rate-equation calculation with different β - factors. Red line represents the best fit to the experimental data (red circles) corresponding to β = 0.27. The fits of 0.1, 0.5, and 1 are also shown for comparison.

# Supplementary Materials

## Note S1. Photonic crystal cavity design

The cavity is formed with three missing air holes (L3 type) within a triangle lattice in a 220 nm silicon membrane with refractive index $n = 3.51$. The second panel of Figure 1b presents the top-view scanning electron microscope image of the PhCC before the fabrication of heterobilayer. The lattice constant of the structure was $a = 296.2$ nm and a radius of the air holes $r = 83.4$ nm. This combination of geometries leads to a relative thickness of $d/a = 0.74$ and relative hole size of $r/a = 0.282$. The two holes nearest to the defect in the line have been shrunk ($r = 59.2$ nm) and shifted outwards by $0.163a$ (48.4 nm) to produce better confinement of the mode, as indicated by the white arrows. Neighboring holes have been reduced by 25% in diameter to decrease their overlap with high field regions (see gray holes in the second panel of Figure 1b). In the calculation, these design concepts achieve a mode volume of $V = 0.68(\lambda/n)^3$ and an effective refractive index $n_{eff}$ of 2.9, a very high value for this class of porous cavity membranes. We optimized the out-of-plane coupling of the cavity mode, and a coupling efficiency of 80% was achieved when using a microscope objective with NA = 0.65.

## Note S2. Raman spectra of monolayer MoS$_2$ and WSe$_2$

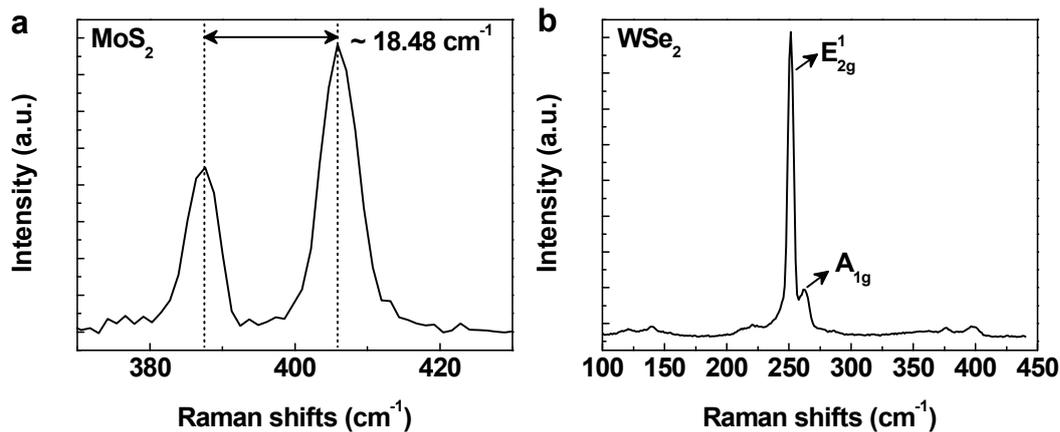

**Fig. S1. Raman spectra of monolayer MoS$_2$ (a) and WSe$_2$ (b).** Raman spectra provides evidence for the monolayer behavior of these two components before the fabrication of the heterobilayer.

## Note S3. Coherence time measurement setup

The coherence time measurement experimental setup is shown in Figure S2. The setup consists of Michelson interferometer and time-correlated photon counting module (not shown in the figure).

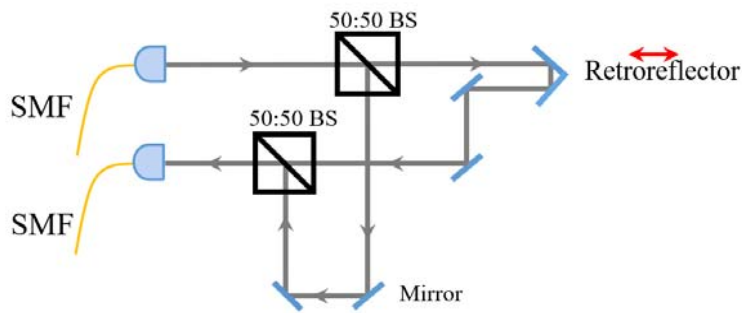

**Fig. S2. Michelson interferometer for coherent length measurement.**

## Note S4. Beta factor fitting

In many cases, it is not needed to utilize the full description based on semiconductor Hamiltonian. Instead, the lasing transition of the semiconductor system can be approximated with a two-level atomic system(*24*). This will result in the classical laser rate equation. This simpler approach has been used successfully to extract the $\beta$ factor of the nanolaser from the L-L curve(*13, 15*).

Using this approach(*15*), the cavity mode intensity $P$ can be expressed as

$$P = \frac{a\Gamma_c R_{ex} - 1/\beta + \sqrt{\left(a\Gamma_c R_{ex} - 1/\beta\right)^2 + 4a\Gamma_c R_{ex}}}{2a}, \quad (S4.1)$$

where $a$ is stimulated emission coefficient, $\Gamma_c$ is equal to the cavity confinement factor divided by the cavity photon decay rate, and $R_{ex}$ is the optical pumping rate. We use this equation to get the theoretical fit of our L-L curve data.

## Note S5. Effect of the lifetime on lasing threshold

The lasing threshold can be obtained by setting $aP = 1$ in equation (S4.1) which results in lasing threshold, $R_{th} = \frac{1 + 1/\beta}{2a\Gamma_c}$. This expression gives a reasonable estimation of the lasing threshold. However, it does not fully describe the dependence of the lasing threshold to the exciton lifetime.

In order to obtain a better description of the effect of lifetime to the lasing threshold we perform the following tasks:
1. Analysis of lifetime dependence of $\beta$ factor
2. Modeling the lasing emission using microscopic theory based on semiconductor Hamiltonian(*24*)

The $\beta$ factor is defined as the ratio between the spontaneous emission that is coupled to the lasing mode and the total spontaneous emission. It depends on the cavity parameter, the emitter lifetime, and the dephasing mechanism. In our case, the exciton spectrum broadening is mainly due to the dephasing caused by inhomogeneous broadening(*25*). In this case, the $\beta$ factor can be treated to be independent of lifetime(*26*). Hence, we can use constant value of $\beta$ in our microscopic equation.

The microscopic theory is obtained by deriving the Heisenberg's equation of motion from the semiconductor Hamiltonian(*24*). The resulting equations is generally not solvable because of the time dependence of the lower order operator is coupled to higher order operator. This is called the operator hierarchy problem(*27*). To handle this problem, the cluster expansion method can be used to truncate the problem to correlation to finite order(*24, 27*).

If we consider the case of off-resonance excitation the photon carrier dynamic can be illustrated as in Fig. S3. Here, $f_e$ ($f_h$) is the electron (hole) occupation number of the lasing transition energy levels while $f_{e2}$ ($f_{h2}$) is the electron (hole) occupation of the off-resonance energy level. The carrier (in our case intralayer exciton) is generated with carrier generation rate $\eta P$ where $P$ is the input power. After it is generated, it can recombine with spontaneous emission rate $t_{SE2}^{-1}$ or relax to the lower energy level with relaxation rate $t_R^{-1}$. The carrier can then recombine with rate equal to $t_{SE}^{-1}$ which results in interlayer exciton spontaneous emission. Additionally, the cavity enhanced photon-carrier interaction will result in stimulated emission and absorption.

If we neglect the Coulomb interaction between the carriers and the exciton-exciton interaction, we will get the following set of differential equations(*27*):

$$\hbar \frac{d}{dt}\langle b^\dagger b\rangle = -2\kappa \langle b^\dagger b\rangle + N\frac{\hbar\beta(\kappa+\Gamma)}{t_{SE}}\langle b^\dagger v^\dagger c\rangle \tag{S5.1}$$

$$\hbar \frac{d}{dt}\langle b^\dagger v^\dagger c\rangle = -(\kappa+\Gamma)\langle b^\dagger v^\dagger c\rangle + f_e f_h - (1-f_e-f_h)\langle b^\dagger b\rangle + \delta\langle b^\dagger b c^\dagger c\rangle - \delta\langle b^\dagger b v^\dagger v\rangle \tag{S5.2}$$

$$\hbar \frac{d}{dt}\delta\langle b^\dagger b c^\dagger c\rangle = -\frac{\hbar\beta(\kappa+\Gamma)}{t_{SE}}\left(\delta\langle b^\dagger b^\dagger b v^\dagger c\rangle + (\langle b^\dagger b\rangle + f_e)\langle b^\dagger v^\dagger c\rangle\right) - 2\kappa\delta\langle b^\dagger b c^\dagger c\rangle \tag{S5.3}$$

$$\hbar \frac{d}{dt}\delta\langle b^\dagger b v^\dagger v\rangle = -\frac{\hbar\beta(\kappa+\Gamma)}{t_{SE}}\left(\delta\langle b^\dagger b^\dagger b v^\dagger c\rangle + (\langle b^\dagger b\rangle + f_h)\langle b^\dagger v^\dagger c\rangle\right) - 2\kappa\delta\langle b^\dagger b v^\dagger v\rangle \tag{S5.5}$$

$$\hbar \frac{d}{dt}\delta\langle b^\dagger b^\dagger bb\rangle = -4\kappa\delta\langle b^\dagger b^\dagger bb\rangle + 2N\frac{\hbar\beta(\kappa+\Gamma)}{t_{SE}}\delta\langle b^\dagger b^\dagger b v^\dagger c\rangle \tag{S5.6}$$

$$\hbar \frac{d}{dt}\delta\langle b^\dagger b^\dagger b v^\dagger c\rangle = -(3\kappa+\Gamma)\delta\langle b^\dagger b^\dagger b v^\dagger c\rangle - \frac{\hbar\beta(\kappa+\Gamma)}{t_{SE}}\langle b^\dagger v^\dagger c\rangle^2 \tag{S5.7}$$

$$-(1-f_e-f_h)\langle b^\dagger b^\dagger bb\rangle + 2\left(f_h\delta\langle b^\dagger b c^\dagger c\rangle - f_e\delta\langle b^\dagger b v^\dagger v\rangle\right)$$

$$\frac{d}{dt}f_e = -\frac{\hbar\beta(\kappa+\Gamma)}{t_{SE}}\langle b^\dagger v^\dagger c\rangle - \frac{(1-\beta)f_e f_h}{t_{SE}} + \frac{(1-f_e)f_{e2}}{t_R} \tag{S5.8}$$

$$\frac{d}{dt}f_h = -\frac{\hbar\beta(\kappa+\Gamma)}{t_{SE}}\langle b^\dagger v^\dagger c\rangle - \frac{(1-\beta)f_e f_h}{t_{SE}} + \frac{(1-f_h)f_{h2}}{t_R} \tag{S5.9}$$

$$\frac{d}{dt}f_{e2} = \eta P(1-f_{e2}-f_{h2}) - \frac{f_{e2}f_{h2}}{t_{SE2}} - \frac{(1-f_e)f_{e2}}{t_R} \tag{S5.10}$$

$$\frac{d}{dt}f_{h2} = \eta P(1-f_{e2}-f_{h2}) - \frac{f_{e2}f_{h2}}{t_{SE2}} - \frac{(1-f_h)f_{h2}}{t_R}, \tag{S5.11}$$

where the $b, v, c$ ($b^\dagger, v^\dagger, c^\dagger$) are the annihilation (creation) operators for cavity mode photon, the hole of interlayer exciton, and the electron of interlayer exciton respectively. The equations are then solved for steady state condition. The constant parameters used are shown in Table S1.

The system reach lasing threshold when the stimulated emission has the same contribution as the spontaneous emission. Using equation (S5.2), this happens when $f_e f_h = -(1-f_e-f_h)\langle b^\dagger b\rangle$. Based on this condition, we can get the lasing threshold as a function of exciton lifetime. The value of the lasing threshold as the lifetime is varied from 0.1 to 1 ns is shown in Fig. S4. As can be seen, the lasing threshold is reduced from 180 µW to 33 µW as the exciton lifetime is increased from 0.1 ns to 1 ns. As the inset of Figure S4 shows the lasing threshold is proportional to the spontaneous emission rate, i.e. it is inversely proportional to the lifetime.

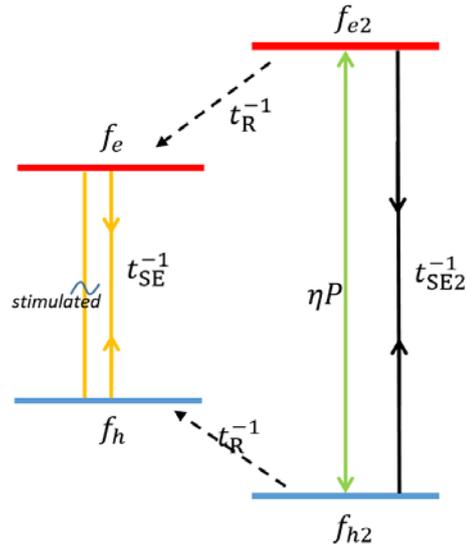

**Fig. S3. Illustration of the light-matter interaction.** Here, $f_e$ ($f_h$) is the electron (hole) occupation number of the on-resonance energy levels while $f_{e2}$ ($f_{h2}$) is the carrier occupation of the off-resonance energy level, $\eta P$ is carrier generation rate, $t_{SE2}^{-1}$ is the spontaneous emission rate of the off-resonance level, $t_R^{-1}$ is the relaxation rate to resonance level, and $t_{SE}^{-1}$ is the interlayer exciton spontaneous emission rate. The cavity enhanced photon-carrier interaction will result in stimulated emission and absorption.

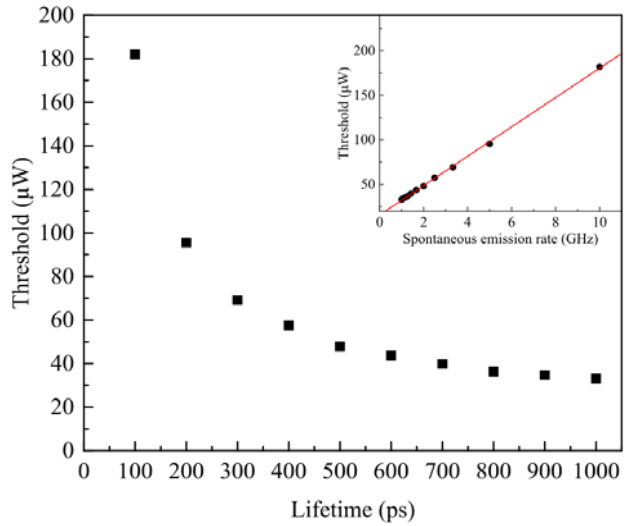

**Fig. S4. Lasing threshold vs. exciton lifetime.** The inset shows a linear fitting of the lasing threshold as a function of spontaneous emission rate.

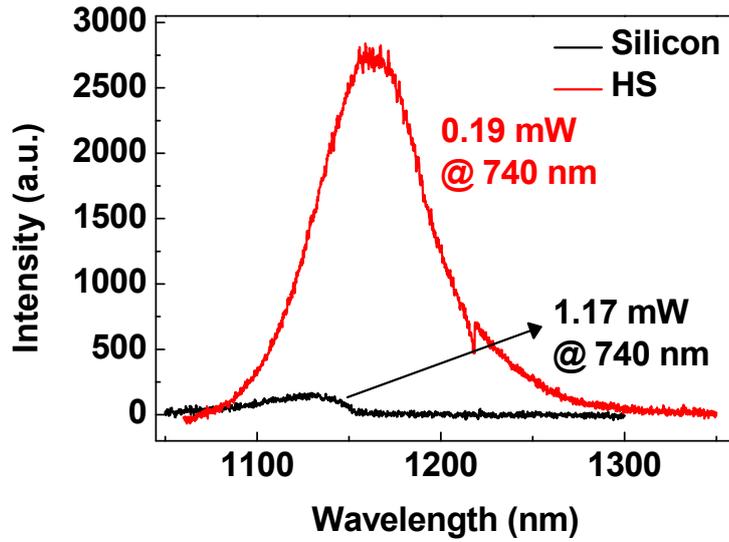

**Fig. S5. Comparison between the PL spectrum from MoS$_2$/WSe$_2$ heterostructure and the PL spectrum from the SOI substrate.** It can be seen that the center wavelength is different and also that the emission from heterostructure is much stronger. This shows that the emission is mainly due to the MoS$_2$/WSe$_2$ interlayer excitonic emission.

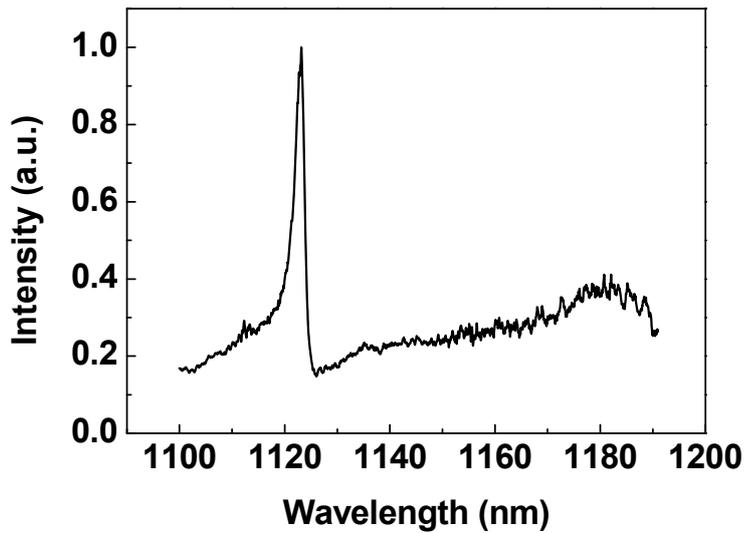

**Figure S6.** Cavity mode of the bare photonic crystal cavity was measured at room temperature using a confocal cross-polarized reflectivity measurement technique(*28*). The fundamental mode can be observed at 1,123 nm.

**Table S1. Definition and values of the parameters used in the theoretical model**

| Symbol | Description | Value |
|---|---|---|
| $\eta$ | Pumping efficiency: [pump rate] = $\eta$ [pump power] | 0.00016 THz/μW |
| $\kappa$ | Cavity loss rate. The value is calculated from the lasing emission linewidth. | 0.55 THz |
| $N$ | Maximum number of carrier. The value is obtained by fitting the L-L curve. | 1400 |
| $\beta$ | $\beta$ factor. The value is obtained by fitting the L-L curve. | 0.16863 |
| $\Gamma$ | Linewidth broadening. The value is calculated from the spontaneous emission linewidth. | 20 THz |
| $t_R$ | Relaxation rate (including the interlayer tunneling) of the carrier | 1 fs |
| $t_{SE2}$ | Spontaneous emission rate of the intralayer exciton into the cavity mode. | 1 ps |